\documentstyle[aps]{revtex}
\begin{document}
\draft
\title{Finite-Sized Plasmas}
\author{Michael Engelhardt\thanks{email: hengel@wicc.weizmann.ac.il} }
\address{Department of Condensed Matter Physics \\
Weizmann Institute of Science \\ Rehovot 76100, Israel }
\date{}
\maketitle

\begin{abstract}
When addressing the thermodynamics of finite-sized systems, one must
specify whether one wants to fix conserved charges to a sharp value
or whether one is content to fix their thermodynamic average. In other
words, contrary to the thermodynamic limit, different statistical
ensembles are not equivalent. When treating the plasma phases of
gauge field theories perturbatively in the canonical ensemble,
unexpected new difficulties arise in comparison with the usual grand
canonical treatment. The purpose of this paper is to expose these
difficulties and show how they can be remedied, thus recovering
a well-defined description of plasmas even in a finite-size,
canonical ensemble setting. For definiteness, a specific model is
considered, namely QCD$_{1+1} $ with SU(2) color; however, the
treatment presented should be applicable also to higher-dimensional
systems with different types of Coulomb interaction.
\end{abstract}

\pacs{PACS number(s): 11.10.Wx, 11.15.-q, 12.38.Cy, 12.38.Mh}

\section{Introduction}
The fundamental interactions governing microscopic behavior in nature
are described by gauge theories. These theories generically give rise
to long-range Coulomb interactions, and for this reason plasma phases
represent an almost universal phenomenon in dense systems, whether one
is considering quarks and gluons in a heavy ion collision or the
electron gas in a metal. Many of the plasmas accessible to experiment
are quite small, not only in the case of the heavy-ion collision
\cite{hwa}, but also e.g. in mesoscopic systems \cite{meso}, which
have become a wealthy source of new phenomena in the past
years. In such a setting, finite-size effects become important, and
they are indeed one of the causes of the high diversity of mesoscopic
physics.

The most obvious effect of a finite configuration space is that the
corresponding momentum space becomes discrete. However, there is also
a slightly more subtle source of new behavior connected with the
ensemble one uses to describe the statistical properties of the system
at hand. Commonly, in the presence of a conserved charge $N$, one uses
the grand canonical ensemble to calculate the partition function,
\begin{equation}
Z_{GC} = \mbox{Tr} \ e^{-\beta (H-\mu N) }
\label{gcens}
\end{equation}
where $\mu $ is the corresponding chemical potential, $\beta $
the inverse temperature, and $H$ the Hamiltonian. One may at
the end of the calculation fix $\mu $ to impose a desired
mean value of $N$. Physically, this corresponds to the system
being able to exchange energy and particles with its surroundings.
In many cases, however, this is an idealization, and the system is
only able to adjust its charge to a limited extent, or not at all.
The prime example is the colliding heavy-ion system, which must remain
an exact color singlet at all times. In such a situation, the
canonical ensemble is more appropriate,
\begin{equation}
Z_C = \int d\alpha \, \mbox{Tr} \
e^{-\beta (H-i\alpha (N-N_0 )/ \beta ) }
\label{cens}
\end{equation}
This restricts the trace to states in which the charge $N$ takes
exactly the value $N_0 $. In the following, $N_0 $ will always be
taken to be zero, without loss of generality.

Whereas for very large systems, (\ref{gcens}) and (\ref{cens}) give
identical results because the fluctuations of $N$ in the grand
canonical ensemble die out as the inverse root of the size, in a
finite system (\ref{gcens}) and (\ref{cens}) are different. Now, at
first glance one may think that the finite size corrections are easy
to obtain if one has, as usual, calculated (\ref{gcens}); simply put
in an imaginary chemical
potential $\mu = i\alpha / \beta $ and do an integration to obtain
(\ref{cens}). Such considerations have been made for ideal gases
e.g. in \cite{turk}-\cite{goren}. A closer look however reveals
that, at imaginary chemical potential, a plasma does not conform to
the recipes conventionally used in its formal treatment. Specifically,
the venerated prescription of \cite{gmbru}, namely summing the
so-called ring diagrams to a logarithm, fails because one runs into
the cut of the logarithm. The main formal advance  of the present
treatment is to show how this difficulty can be remedied. Thus it will
be possible to regain a well-defined description of the plasma
even in a finite-size, canonical ensemble setting.

The specific model to be considered here is QCD in one space dimension
in the high temperature or high baryon density perturbative regime
\cite{mcl}. It must be emphasized however that the formal developments
presented are quite general and are straightforwardly applied also to
higher-dimensional systems with any type of Coulombic interaction. A
one-dimensional model is chosen because in this setting, the infrared
divergences appear most clearly, uncluttered by the usual ultraviolet
problems. The results however are expected to enjoy a wide range of
applicability. The goal of the treatment is to evaluate the
contribution of the plasmon effect to the thermodynamics of a finite
system, exhibiting especially the differences in behavior away from
the thermodynamic limit depending on which ensemble is used. Of
course, the size of the system is still taken to be large enough to
permit a physical interpretation as a plasma, i.e. if $g$ is the
coupling constant in the relevant Lagrangian and $L$ is the length of
the system, then
$gL$ should still be large\footnote{Note that $gL$ is dimensionless
in one space dimension.}. Otherwise, the concept of screening the
Coulomb potential becomes meaningless and the perturbative expansion
of the theory is in powers of $g^2 L^2 $. Physically, large $gL$ means
that the plasmons, whose size is controlled by $1/g$ in one space
dimension, still fit well into the length $L$. Correspondingly, in
three space dimensions, the plasmon size behaves \cite{lvw} like
$1/gT$ and thus the relevant
dimensionless parameter which must remain sizeable would be
$gTL$. Only under such conditions is it meaningful to consider the
usual plasmon contribution to the thermodynamic potential and evaluate
its behavior as $1/gL$ is increased away from zero.

\section{The Model}
The axial gauge Hamiltonian of QCD$_{1+1} $ with two colors reads in
momentum space
\begin{equation}
H=\int dp \, \chi^{\dagger } (p) (p\gamma_{5} +m\gamma_{0} ) \chi (p)
+ \frac{g^2 }{4\pi } \int \frac{dq}{q^2 } j^a (q) j^a (-q)
\label{ham}
\end{equation}
where the fermion fields $\chi $ carry Dirac and color indices and
the SU(2) currents $j^a $ are
\begin{equation}
j^a (q) = \int dp \, \chi^{\dagger } (p)
\frac{\sigma^{a} }{2} \chi (p+q)
\end{equation}
with the Pauli matrices $\sigma^{a} $. For later use it is also useful
to introduce the U(1) currents
\begin{equation}
j^0 (q) = \frac{1}{2} \int dp \, \chi^{\dagger } (p) \chi (p+q)
\end{equation}
The Hamiltonian (\ref{ham}) conserves the U(1) charge $j^0 (0)$,
corresponding to the baryon number, and the SU(2) color charges
$j^a (0)$. For definiteness, it will be assumed that the plasma may
exchange baryon number with its surroundings, but not color charge.
This will illustrate all the possible complications; all alternative
scenarios can be treated along similar lines. In particular, it is
straightforward e.g. to specialize to QED in the canonical ensemble.
Thus, the goal is to evaluate the partition function
\begin{equation}
Z=\int_{SU(2)} dG(\vec{\alpha } ) \mbox{ Tr } \exp \left[
-\beta (H-\mu j^0 (0) - i\alpha^{a} j^a (0) / \beta ) \right]
\end{equation}
where $dG(\vec{\alpha } )$ denotes the Haar measure of SU(2)
\cite{turk}-\cite{goren}. Generically,
a further simplification is possible at this point. The trace is
invariant under a global unitary transformation of the fermion fields,
\begin{equation}
\chi \rightarrow \chi^{\prime } = U\chi
\end{equation}
and also the Hamiltonian and the baryon number remain unchanged under
this transformation (this is a global residual gauge invariance still
present after gauge fixing). Therefore, one may choose to diagonalize
$\alpha^{a} \sigma^{a} $,
\begin{equation}
U^{\dagger } \alpha^{a} \sigma^{a} U = \sigma^{3}
\sqrt{ (\alpha^{1} )^2 + (\alpha^{2} )^2 + (\alpha^{3} )^2 }
\equiv \sigma^{3} \alpha
\label{diagal}
\end{equation}
i.e. the partition function depends only on the length of
$\vec{\alpha } $, not its direction in color space,
\begin{equation}
Z= \frac{1}{2\pi } \int_{-2\pi }^{2\pi } d\alpha \, \sin^{2}
\frac{\alpha }{2} \mbox{ Tr } \exp \left[
-\beta (H-\mu j^0 (0) - i\alpha j^3 (0) /\beta ) \right]
\equiv \frac{1}{2\pi } \int_{-2\pi }^{2\pi } d\alpha \, \sin^{2}
\frac{\alpha }{2} Z(\alpha )
\label{zust}
\end{equation}
In general, only the (imaginary) chemical potentials associated with
the Cartan subalgebra of the gauge group are relevant. In addition to
the symmetry exploited above, there still remains a discrete
translational symmetry in $\alpha $ and also a parity symmetry
corresponding to the freedom of permuting the eigenvalues in the
diagonalized matrix in (\ref{diagal}).

In
the following, (\ref{zust}) will be referred to as the ``canonical''
partition function (even though the U(1) charge is treated grand
canonically). On the other hand, treating also the SU(2) color charges
grand canonically (this will be referred to as the ``grand canonical''
case) corresponds simply to dropping the average over $\alpha $ in
(\ref{zust}) and setting $Z=Z(\alpha =0)$.

It
should be mentioned that in the specific setting of QCD$_{1+1} $, the
Hamiltonian (\ref{ham}) is only valid in the color singlet sector. In
general, the Hamiltonian of QCD$_{1+1} $ contains additional couplings
to residual gauge degrees of freedom \cite{elim}, the dynamics of
which eliminate all non-singlet states from the spectrum. Thus, the
partition function of QCD$_{1+1} $ strictly speaking contains only
singlet states
already by virtue of the dynamics. Of course it is fully equivalent to
use the color singlet Hamiltonian (\ref{ham}) and impose the color
singlet constraint by hand as in (\ref{zust}). On the other hand, for
purposes of comparison below also the grand canonical partition
function corresponding to the Hamiltonian (\ref{ham}) will be
considered. Strictly
speaking, this is not a physically meaningful alternative. Therefore,
from the point of view of QCD$_{1+1} $ what is being compared is
a physically correct calculation (namely, the canonical one) with the
idealized grand canonical one which is often adopted in the literature
due to technical convenience. Of course, in other settings, such as
electrodynamical plasmas in three space dimensions, both alternatives
may be physically meaningful, depending on the laboratory conditions.

\section{Ideal Gas}
Momentarily neglecting the Coulomb interaction, one can now
immediately read off the ideal gas partition function \cite{kapus},
\begin{eqnarray}
\ln Z_0 (\alpha ) &=& \frac{L}{2\pi } \int dp \sum_{i} \left[
\ln (1+e^{-\beta (\epsilon_{p} +\mu /2 +i\sigma^{3}_{ii}
\alpha /2\beta )} )
+\ln (1+e^{-\beta (\epsilon_{p} -\mu /2 -i\sigma^{3}_{ii}
\alpha /2\beta )} )
\right] \\ &=& \frac{L}{2\pi } \int dp \, \left[
\ln (1+2\cos (\alpha /2) e^{-\beta (\epsilon_{p} +\mu /2)} +
e^{-2\beta (\epsilon_{p} +\mu /2)} ) \right. \nonumber \\ & &
\ \ \ \ \ \ \ \ \ \ \ \ \ \ \ \ \left. +
\ln (1+2\cos (\alpha /2) e^{-\beta (\epsilon_{p} -\mu /2)} +
e^{-2\beta (\epsilon_{p} -\mu /2)} ) \right]
\label{frei}
\end{eqnarray}
where $\epsilon_{p} = \sqrt{p^2 + m^2 } $. In some special cases, this
can be evaluated explicitely as long as one still assumes the momenta
to be continuous. E.g. in the classical nonrelativistic limit
$m\gg m-\mu /2 \gg T$ one obtains
\begin{equation}
\ln Z_0 (\alpha ) = \frac{L}{\pi } \cos (\alpha /2)
\sqrt{\frac{2m\pi }{\beta } } e^{-\beta (m-\mu /2 ) }
\label{frklnr}
\end{equation}
whereas in the limit $m=\mu =0$ one has \cite{grads}
\begin{equation}
\ln Z_0 (\alpha ) = L\left( \frac{\pi }{3\beta } -
\frac{\alpha^{2} }{4\pi \beta } \right)
\label{frm0}
\end{equation}

Taking now into account the interaction perturbatively, assume that
$\ln Z(\alpha ) $ has been calculated to some order in $g$ (note that
one usually calculates directly $\ln Z_I (\alpha ) $ by considering
only connected Feynman diagrams),
\begin{equation}
\ln Z(\alpha ) = \ln Z_0 (\alpha ) + \ln Z_I (\alpha )
\end{equation}
According to (\ref{zust}), this must now be averaged over $\alpha $,
\begin{eqnarray}
\ln Z &=& \ln \frac{1}{2\pi } \int d\alpha \, \sin^{2}
\frac{\alpha }{2}
Z_0 (\alpha ) Z_I (\alpha ) \\
&=& \ln \frac{1}{2\pi } \int d\alpha \, \sin^{2} \frac{\alpha }{2}
Z_0 (\alpha ) + \ln \frac{\int d\alpha \, \sin^{2} (\alpha /2)
Z_0 (\alpha ) Z_I (\alpha ) }{\int d\alpha \, \sin^{2} (\alpha /2)
Z_0 (\alpha ) }
\label{aver}
\end{eqnarray}
Thus $Z_0 (\alpha ) $, apart from giving the ideal gas contribution in
the first term of (\ref{aver}), also acts as a measure for averaging
$Z_I (\alpha ) $. Now, inspection of (\ref{frklnr}) and (\ref{frm0}),
or in general (\ref{frei}), reveals that
$Z_0 (\alpha ) $ falls off as a Gaussian with a width of $1/\sqrt{L} $
in $\alpha $ for large $L$. Therefore, in the thermodynamic limit,
only the vicinity of $\alpha =0$ contributes to $\ln Z $. Thus one
recovers the well-known result that the canonical and grand canonical
ensembles are equivalent for large systems, or, formulated in another
way, color singlet constraints become irrelevant in the thermodynamic
limit.

For finite
$L$, on the other hand, the two ensembles differ. Whereas the grand
canonical ideal gas partition function is obtained simply by setting
$\alpha =0$ in (\ref{frei}), (\ref{frklnr}), or (\ref{frm0}), in the
canonical ensemble one must average over $\alpha $. In special cases,
this can be accomplished analytically. In the classical nonrelativistic
limit, the average over (\ref{frklnr}) can be carried out to give
\begin{equation}
\ln Z_0 = \ln \left( 2I_1 (x) /x \right) \ \ \ \ \mbox{with} \ \ \ \
x=\frac{L}{\pi } \sqrt{\frac{2m\pi }{\beta } } e^{-\beta (m-\mu /2) }
\label{avklnr}
\end{equation}
where $I_1 (x)$ is a Bessel function of imaginary argument
\cite{grads}.
In the case $m=\mu =0$ one obtains using (\ref{frm0})
\begin{equation}
\ln Z_0 = \frac{L\pi }{3\beta } + \ln \left[ \frac{1}{4\sqrt{ \pi x} }
\left( \mbox{erf} (2\pi \sqrt{x} ) - e^{-\frac{1}{4x} } \left(
\mbox{Re erf} (2\pi \sqrt{x} + \frac{i}{2\sqrt{x} } )
-\mbox{Re erf} (\frac{i}{2\sqrt{x} } ) \right) \right) \right]
\label{avm0}
\end{equation}
where erf denotes the error function and
$x=L/(4\pi \beta )$. The comparison between the ensembles is
illustrated for the ideal gas in figures (\ref{fig1}) and (\ref{fig2}).
Also, to give an idea how the effect of discretizing the momenta compares
with the effect of varying the ensemble, the free partition function
has been calculated using antiperiodic boundary conditions for the
fermion fields, i.e.
\begin{equation}
p \rightarrow p_n = \frac{(2n+1)\pi }{L} , \ \ \ \ \int dp \rightarrow
\frac{2\pi }{L} \sum_{p_n }
\end{equation}
as well as periodic boundary conditions,
\begin{equation}
p \rightarrow p_n = \frac{2n\pi }{L} , \ \ \ \ \int dp \rightarrow
\frac{2\pi }{L} \sum_{p_n }
\end{equation}
Evidently
both figures, though stemming from vastly different regions of the
phase diagram, display a dominance of the effect due to varying choice
of ensemble except at very low $L$. Formally, what happens is that the
difference between an integral and the corresponding Riemann sum
behaves as $1/L$, as can be inferred from the Euler-McLaurin summation
formula \cite{abr}. Thus, merely discretizing the momenta leads to
an expansion of the form $\ln Z_0 \sim O(L) + O(1) + \ldots $ for the
logarithm of the partition function. On the
other hand, the averaging procedure involved in doing a canonical
calculation in general introduces power corrections in $L$ into the
partition function $Z_0 $, as can be explicitely observed in
(\ref{avklnr}) and (\ref{avm0}). This leads to an expansion of the
form $\ln Z_0 \sim O(L) + O(\ln L) + \ldots $ upon taking the
logarithm. At low $L$, the question which effect dominates
depends on the details of the dynamics, as evidenced in figures
(\ref{fig1}) and (\ref{fig2}). Whereas in figure (\ref{fig2}),
the discretization effects are negligible even at $LT=1$, in the
ultrarelativistic case the discretization effects become comparable to
the effect of varying the ensemble at sufficiently low $L$. The effect
of using a canonical ensemble is reinforced by using antiperiodic
boundary conditions and weakened by using periodic ones. A similar
comparison will be possible for the plasmon contribution calculated in
the next section.

\section{Plasmon contribution}
This
section is concerned with the main object of the present treatment,
the perturbative evaluation of the plasmon contribution to the
thermodynamic potential. Whereas this effect arises only at order
$O(g^3 )$ in gauge theories in three space dimensions \cite{kapus},
it already contributes at order $O(g)$ in one space dimension due to
the different volume element in momentum space. Thus it represents the
dominant perturbative effect here.

As a side remark, note that in QCD$_{1+1} $, this is in fact the only
term of the perturbative expansion which can be obtained in terms of
Feynman diagrams. Already in the next order of perturbation theory,
the Feynman expansion breaks down due to irremediable ambiguities
introduced by the infrared divergences \cite{thesis}. Formally, one
can give convergent sums over selected subclasses of diagrams such
that virtually arbitrary fractional powers (greater than one) of the
coupling constant $g$ are generated.
On the other hand, it is known e.g. in the classical
nonrelativistic limit of QCD$_{1+1} $ \cite{let} that expanding the
exact equation of state in the coupling constant yields a series made
up exclusively of integer powers of $g$. Similar infrared problems
arise in QCD$_{3+1} $ at order $O(g^6 )$ \cite{linde}. In the case of
QED, the infrared divergences are not as severe and the Feynman
expansion can be carried out at least to some higher order in $g$.

The perturbative expansion of the expression (\ref{zust}) for the
partition function may be obtained by standard methods \cite{kapus}
\cite{fuw}. The corresponding Feynman rules can be read off from
(\ref{zust}) and (\ref{ham}) as follows:
\begin{itemize}
\item
Fermion
propagator with momentum $p$, color $i$ and Matsubara frequency
$\omega_{s} = (2s+1)\pi /\beta $:
\begin{equation}
K_i (p,s) = \frac{i}{\beta } \
\frac{\omega_{s} -i\mu /2 + \sigma^{3}_{ii} \alpha /2\beta
+i\gamma_{5} p -i\gamma_{0} m}{(\omega_{s} -i\mu /2
+\sigma^{3}_{ii} \alpha /2\beta )^2 +p^2 +m^2 }
\end{equation}
\item
Coulomb interaction with momentum transfer $q$ and energy transfer
$2\pi r/\beta $ (cf. figure (\ref{vertex})):
\begin{equation}
V_{ij,kl} (q,r) = -\frac{g^2 \beta }{16 \pi } \sum_{a}
\frac{\sigma^{a}_{ji} \sigma^{a}_{lk} }{q^2 }
\end{equation}
\end{itemize}
All internal momenta, frequencies, color and Dirac indices must be
integrated or summed over, respectively; fermion loops are associated
with an additional factor $(-1)$, and diagrams with $N$
Coulomb interactions have an overall factor
$1/N! $. Every connected part of a diagram in addition receives a
factor $L/2\pi $; note that these rules correspond to already
having used momentum conservation at all vertices. By considering only
all connected diagrams, one directly obtains the logarithm of the
partition function, an extensive quantity in the infinite volume
limit.

The plasmon contribution $\ln Z_R $ is obtained by summing the ring
diagrams (cf. figure (\ref{ringsum})). Physically, this is interpreted
as a screening of the Coulomb interaction by the production of
fermion-antifermion pairs; thus the infrared divergences resulting from
the long-range character of the interaction are dampened. Since there
are $2^{N-1} (N-1)! $ ways of combining $N$ vertices to a ring, the
Feynman rules give
\begin{equation}
\ln Z_R (\alpha ) = \frac{L}{4\pi } \sum_{r} \int dq
\mbox{ Tr}^{color}
\sum_{N=1}^{\infty } \frac{1}{N}
\left( \frac{\Pi (r,q=0)}{q^2 } \right)^{N}
\label{rsum}
\end{equation}
where $r$ is the integer labeling the frequency flow around the ring,
$q$ is the momentum flow around the ring, and the polarization $\Pi $
is a matrix in color space,
\begin{equation}
\Pi^{a\bar{a} } (r,0) = -\frac{g^2 }{4\pi \beta } \sum_{i,j}
\int dp \sum_{s} \sigma_{ji}^{a} \sigma_{ij}^{\bar{a} }
\frac{(\omega_{s} -i\frac{\mu }{2} + \frac{\alpha }{2\beta }
\sigma^{3}_{ii} )(\omega_{s+r} -i\frac{\mu }{2} +
\frac{\alpha }{2\beta }
\sigma^{3}_{jj} ) -\epsilon_{p}^{2} }{[(\omega_{s} -i\frac{\mu }{2}
+\frac{\alpha }{2\beta } \sigma^{3}_{ii} )^2 +\epsilon_{p}^{2} ]
[(\omega_{s+r} -i\frac{\mu }{2} +\frac{\alpha }{2\beta }
\sigma^{3}_{jj} )^2 +\epsilon_{p}^{2} ]}
\label{polariz}
\end{equation}
Note that in (\ref{rsum}), only the value of the polarization at
$q=0$ is taken. Higher orders in the Taylor expansion of $\Pi $ around
$q=0$ dampen the infrared divergence and thus ultimately lead only to
terms of higher order in the perturbative expansion
in $g$ \cite{kapus}.
The sum over $s$ in (\ref{polariz}) is readily carried out by standard
methods \cite{fuw},
\begin{eqnarray}
\Pi^{a\bar{a} } (r,0) &=& g^2 \sum_{i,j}
\sigma^{a}_{ji} \sigma^{\bar{a} }_{ij}
\frac{\sin \frac{\alpha }{4}
(\sigma^{3}_{jj} -\sigma^{3}_{ii} )}{\pi r
-\frac{\alpha }{4} (\sigma^{3}_{jj} -\sigma^{3}_{ii} ) } G_{ji} \\
G_{ji} (\alpha ) &=& \frac{\beta }{32\pi } \int dp \left[
\frac{1}{\cosh \frac{\beta }{2} (\epsilon_{p} + \frac{\mu }{2}
+i\frac{\alpha }{2\beta } \sigma^{3}_{jj} )
\cosh \frac{\beta }{2} (\epsilon_{p} + \frac{\mu }{2}
+i\frac{\alpha }{2\beta } \sigma^{3}_{ii} ) } \right. \nonumber \\
& & \ \ \ \ \ \ \ \ \ \ \ \ \ \ \ \ \ \ \ \ \left.
+\frac{1}{\cosh \frac{\beta }{2} (\epsilon_{p} -\frac{\mu }{2}
-i\frac{\alpha }{2\beta } \sigma^{3}_{jj} )
\cosh \frac{\beta }{2} (\epsilon_{p} - \frac{\mu }{2}
-i\frac{\alpha }{2\beta } \sigma^{3}_{ii} ) } \right]
\end{eqnarray}
Note
$G_{22} = G_{11}^{*} $ and $G_{21} =G_{12} \ge 0$. Furthermore, it is
advantageous to diagonalize $\Pi $ in color space, since (\ref{rsum})
calls for the trace over arbitrary powers of $\Pi $. One arrives at
the eigenvalues
\begin{eqnarray}
\Pi_{1} (r) &=& 2 g^2 \mbox{ Re} \, G_{11} (\alpha ) \left.
\frac{\sin \frac{\epsilon }{2} }{\pi r -\frac{\epsilon }{2} }
\right|_{\epsilon \rightarrow 0} \label{pi1} \\
\Pi_{2} (r) &=& 2 g^2 G_{12} (\alpha )
\frac{\sin \frac{\alpha }{2} }{\pi r -\frac{\alpha }{2} } \label{pi2} \\
\Pi_{3} (r) &=& 2 g^2 G_{12} (\alpha )
\frac{\sin \frac{\alpha }{2} }{-\pi r -\frac{\alpha }{2} } \label{pi3}
\end{eqnarray}
Why $\Pi_{1} $ is left in this form will become clear below. Note that
in the case of electrodynamics, the only term which appears is of the
form of $\Pi_{1} $; the SU(2) case thus exhibits additional structures
compared with the U(1) case.

Cast as above, it becomes very transparent where the situation at
finite imaginary chemical potential differs from the usual one.
According to (\ref{rsum}), one must now calculate
\begin{equation}
\ln Z_R (\alpha ) =\frac{L}{4\pi } \int dq \sum_{i} \sum_{r}
\sum_{N=1}^{\infty } \frac{1}{N}
\left( \frac{\Pi_{i} }{q^2 } \right)^{N}
\label{ralfsum}
\end{equation}
Usually, i.e. for $\alpha =0$, one would observe that
(\ref{pi1})-(\ref{pi3}) only give contributions for $r=0$ and are
negative there. Thus, the sum over $r$ contains only one term, one may
carry out the sum over $N$ to obtain a logarithm, and finally the
(now infrared-regular) $q$-integral may be done \cite{gmbru}
\cite{fuw}.

In the more general case considered here, it is not meaningful to sum to
a logarithm, since $\Pi_{2} $ and $\Pi_{3} $ may have positive values
depending on the value of $r$; also $\Pi_{1} $, though only giving a
contribution for $r=0$, may become positive because Re$\, G_{11} $
turns to negative values at some finite $\alpha $. However, as soon as
the $\Pi_{i} $ become positive, one only has to go to sufficiently
low $q$ to run into the cut of the logarithm to which one may have
naively summed (\ref{ralfsum}).

There does not seem to be a physical interpretation for this failure.
One is simply not doing a good job of defining what the initially
ill-defined sum over $N$ in (\ref{ralfsum}) should be; the usual
summation to a logarithm replaces one ill-defined quantity with another
one. There is no physical principle which may e.g. tell one how to
circumvent the cut in the logarithm by giving a small imaginary part
to the momentum $q$. It must be emphasized that this problem has nothing
to do with the question of how the infrared singularity in the
Coulomb interaction (cf. (\ref{ham})) should be defined. It already
occurs for finite, sufficiently large $L$, where the momenta are
discrete and where the Coulomb interaction consequently is completely
well-defined (at least in the color-singlet sector considered here).
The difficulty thus seems to be more related to the
asymptotic nature of the perturbative expansion itself.

However, one may indeed do a better job of defining (\ref{ralfsum})
by proceeding in a different way, namely by exploiting the structure
of (\ref{pi1})-(\ref{pi3}) in $r$; this is the reason why $\Pi_{1} $
was not explicitely simplified to $\Pi_{1} =-2\delta_{r0}
\mbox{ Re} \, G_{11} $. The key observation is that
\begin{equation}
\sum_{r} \frac{1}{(\pi r -\frac{\alpha }{2} )^N } =
\frac{1}{2} \sum_{k\neq 0} \frac{1}{|k|} e^{ik\alpha }
\frac{1}{(N-1)!} (2ik)^N
\label{trick}
\end{equation}
(this identity is derived in the appendix). In this way, one generates
a crucial additional factor $1/(N-1)!$ which turns the sum over $N$
in (\ref{ralfsum}) into an exponential. Therefore one finally obtains
\begin{equation}
\ln Z_R (\alpha ) \! =\! \frac{L}{8\pi } \sum_{k\neq 0} \frac{1}{|k|}
\int \! dq \left[ e^{ik\epsilon } (e^{4ikg^2 \sin (\frac{\epsilon }{2} )
\mbox{\scriptsize Re } G_{11} (\alpha ) /q^2 } \! -1)
\right|_{\epsilon \rightarrow 0}
\left.
+2e^{ik\alpha } (e^{4ikg^2 \sin (\frac{\alpha }{2} ) G_{12}
(\alpha ) /q^2 } \! -1)
\right]
\label{fundex}
\end{equation}
This expression is completely well-defined. It represents the main
improvement of the present treatment over more conventional
ones. Note that the procedure above is reminiscent of the Borel
summation method \cite{fisch}, where one introduces a factor
$1=(k!)^{-1} \int_{0}^{\infty } dt\, e^{-t} t^k $ to generate better
convergence of a series. The price one pays is the introduction of an
additional integration. Here, on the other hand, the
convergence-improving factor is generated
simply by rearranging the already present $r$-summation.
A certain price is implied by the poor numerical convergence
of the $k$-series as it stands. How this is best handled will be
commented upon further below.

In
general, if the momenta are discrete, one must proceed from equation
(\ref{fundex}) numerically. As long as the momenta are
continuous, one may do the integral over $q$ \cite{grads} (due to the
reflection symmetry of (\ref{fundex}) in $\alpha $ and $\epsilon $
around zero, these parameters will be taken positive in the
following):
\begin{eqnarray}
\ln Z_R (\alpha ) &=& \frac{gL}{\sqrt{8\pi } } \sum_{k\neq 0}
\frac{1}{|k|} \left[ e^{ik\epsilon }
\sqrt{|k| \sin \frac{\epsilon }{2}
|\mbox{Re} \, G_{11} (\alpha ) | } \,
(-1+ i \, \mbox{sgn} (k) \, \mbox{sgn}
(\mbox{Re} \, G_{11} (\alpha ) ) ) \right|_{\epsilon \rightarrow 0}
\nonumber \\ & & \ \ \ \ \ \ \ \ \ \ \ \ \ \ \ \ \ \ \ \ \ \ \ \ \ \
\left.
+2e^{ik\alpha } \sqrt{|k| \sin \frac{\alpha }{2} G_{12} (\alpha ) } \,
(-1 +i \, \mbox{sgn} (k) ) \right]
\label{qcon}
\end{eqnarray}
Furthermore, in the case $\alpha \rightarrow 0$, one may use
\begin{equation}
\lim_{\alpha \rightarrow 0} \sum_{k\neq 0} \frac{1}{\sqrt{|k|} }
e^{ik\alpha } \sqrt{\sin \frac{\alpha }{2} } \,
(-1 +i \, \mbox{sgn} (k) )
=-2\int_{0}^{\infty } dx \, \frac{\sin (x+\pi /4)}{\sqrt{x} }
=-2\sqrt{\pi }
\end{equation}
and in this way verify that in a grand canonical treatment, or in a
canonical treatment in the thermodynamic limit, one reproduces
precisely the result of the usual prescription outlined directly after
equation (\ref{ralfsum}). As a side remark, note that taking
additionally the classical nonrelativistic limit allows to evaluate
$G$ explicitely to give
\begin{equation}
\ln Z_R = -\frac{3}{4} gL \left( \frac{2m\beta }{\pi } \right)^{1/4}
e^{-\frac{\beta }{2} (m-\mu /2) }
\end{equation}
Combining this with the corresponding ideal gas result, determining
$\rho (\mu ,T) = \partial P /\partial \mu |_{T} $, and solving for
$\mu $ to obtain the equation of state yields
\begin{equation}
P=\rho T \left( 2-\frac{3\sqrt{2} }{8}
\frac{g}{\sqrt{\rho T} } + \ldots \right)
\end{equation}
This
result was obtained in \cite{let} by a completely different method and
thus provides a nice check on the present calculation. Similarly, one
obtains in the case $m=0$ (still in the thermodynamic limit)
\begin{equation}
\ln Z_R = -\frac{3}{4} \sqrt{\frac{2}{\pi } } gL
\end{equation}
Some further results for finite $\alpha $ are collected in the
appendix.

With the help
of the general formula (\ref{fundex}), one may now again perform
a comparison of the different ensembles. Note that in practice,
it is advantageous to perform the $\alpha $-averaging over
(\ref{fundex})
before doing the $k$-sum. Then the $k$-sum converges quite well above
$k\sim \sqrt{L} $ for large $L$, as can be inferred from stationary
phase considerations. Of course, adopting this procedure forces one to
treat rapidly oscillating $\alpha $-integrands for large $k$, but this
is numerically the lesser evil. Also, as before, allowance was made
for a discretization of the momenta. Note that discretizing
the momentum \underline{transfer} $q$ in (\ref{fundex}), which can be
interpreted as the plasmon momentum, always means taking
\begin{equation}
q \rightarrow q_n = \frac{2\pi n}{L} , \ \ \ \ \int dq \rightarrow
\frac{2\pi }{L} \sum_{q_n \neq 0}
\end{equation}
regardless
of whether one is considering antiperiodic or periodic boundary
conditions for the fermion fields. Note also that the term $q=0$ is
excluded from the sum. This is due to the fact that this term originates
directly from the $q=0$ term in the original Coulomb interaction
(cf. (\ref{ham})), proportional to $j^a (0) j^a (0) /q^2 $. However,
$j^a (0) $ is being constrained to be zero and therefore this term
may be excluded from the Coulomb interaction. Physically, this simply
takes account of the fact that for a neutral system, there is no
physical content in a fluctuation translating all charges in the
system uniformly.

The comparison between the different cases is illustrated in figures
(\ref{fig5}) and (\ref{fig6}). The pictures are strikingly similar,
even though taken from vastly different regions of the phase diagram.
Again, at very high $L$, the effect of varying the ensemble dominates.
However, in contradistinction to the ideal gas term, the effect of
discretizing the momenta takes over already at quite high
$L$. This does not come as a complete surprise. The simple argument
used in the discussion of the ideal gas term, that the Euler-McLaurin
series implies an expansion purely in powers of $1/L$ for the logarithm
of the partition function if one discretizes the momenta, is not valid
for the integral over the plasmon momentum $q$ due to the singular
nature of the integrand in (\ref{fundex}) around $q=0$. Thus, there is
no simple argument anymore which would allow one to predict
which effect dominates for large
$L$. That the strong effect of momentum discretization here is indeed
due to the plasmon momentum sum, as opposed to the fermion momenta
contained in $\ln Z_0 $ and the quantity $G$, is corroborated by
the observation that figures (\ref{fig5}) and  (\ref{fig6}) remain
identical regardless of whether one uses periodic or antiperiodic
boundary conditions for the fermions. Only at very low $L$, not displayed
in the figures anymore, could one observe a dependence on the boundary
conditions. This happens at such low $L$ that the entire concept of
treating the system as a plasma becomes invalid. Note that this is also
true for the ideal gas contribution depicted in figures (\ref{fig1})
and (\ref{fig2}). Below $L/\beta \approx 10$, it is impossible to
simultaneously maintain $gL \gg 1$ and $g\beta \ll 1$.

Thus, the only relevant finite-size effects in the plasmon
contribution stem from the use of a canonical ensemble and from
discretizing the plasmon momentum. The two effects are roughly
comparable in magnitude, with the ensemble effect dominating at
very high $L$.

\section{Discussion}
The present investigation has focussed on the different finite-size
effects influencing the thermodynamics of a plasma. Special emphasis
was placed on the evaluation of the plasmon contribution in a canonical
ensemble as opposed to the usual grand canonical treatment. Here,
it turned out that the usual resummation prescription for the ring
diagrams is too naive, and it was shown how the resulting ambiguity
may be remedied via equation (\ref{trick}). This latter trick, which
in view of its derivation (\ref{tder1})-(\ref{tdern}) may be
suggestively termed ``Fourier regularization'', may be useful in
quite diverse contexts as a means of properly defining ill-defined
series. In the present context, it should be noted that the
nontrivial manipulations purely involve the imaginary time direction
or the corresponding frequency sums. Thus, while the present treatment
focussed on a one-dimensional model, all manipulations are equally
applicable for the analogous problem in an arbitrary number of
space dimensions. Also, the versatility of the method is evidenced by
its application to the more complicated SU(2) Coulomb interaction
as opposed to the U(1) one.

Using
these formal developments, the partition function of QCD$_{1+1} $
for SU(2) color was calculated to order $O(g) $, exhibiting the effect
of varying the ensemble used and of discretizing the momenta.
The dominant finite-size effects were identified to be the ones
resulting from using a canonical ensemble and from discretizing the
plasmon momentum. The effect of discretizing the fermion momenta in
comparison was found to be negligible for interesting values of $L$.
Of the two dominant effects mentioned above, the ensemble effect is
strongest at very high $L$.

Finally,
it should be mentioned that there is an alternative interpretation
of the imaginary chemical potential $\alpha $ over which one averages
in the canonical ensemble: The configuration space of the model
considered here is a cylinder,
for finite $L$ possibly closed to a torus depending on the spatial
boundary conditions used. Moving on the cylinder parallel to its axis
corresponds to moving in the space direction, moving around the
cylinder corresponds to moving in the imaginary time direction. The
parameter $\alpha $ can be interpreted as an (imaginary) background
chromoelectric potential as would be produced by a hypothetical solenoid
coinciding with the axis of the cylinder. This background potential
introduces a phase into Green's functions which lead around the cylinder,
or in more physical terms, the $\alpha $-dependence corresponds to the
Aharonov-Bohm effect induced by the solenoid on the thermodynamics.
A similar situation was investigated recently in connection with
the statistical properties of the Gross-Neveu model on a ring
\cite{bernd}. One of the interesting results of this latter treatment
was the observation of non-analytic behavior of the thermodynamical
observables as a function of the Aharonov-Bohm phase. Disregarding for
the moment that in the present calculation, the parameter $\alpha $
is in the end an integration variable, similar behavior is found here
in quantities before $\alpha $-averaging, e.g. in the first term in
the square brackets of the expression (\ref{disc}) for
$\ln Z_R (\alpha )$. It may be worthwhile to investigate
whether there is physical meaning contained in this singularity.

\subsection*{Acknowledgements}
The author wishes to thank Y.Gefen and A.Ruckenstein for
encouragement. This work was supported by a MINERVA fellowship.

\begin{appendix}
\section*{Some useful Derivations and Formulas}
Identity (\ref{trick}) can be seen in the following way:
\begin{eqnarray}
\sum_{r} \frac{1}{(\pi r -\frac{\alpha }{2} )^N } &=&
\frac{2^{N-1} }{(N-1)!} \left( \frac{d}{d\alpha } \right)^{N-1}
\sum_{r} \frac{1}{\pi r -\frac{\alpha }{2} } \label{tder1} \\
&=& -\frac{2^{N-1} }{(N-1)!} \left( \frac{d}{d\alpha } \right)^{N-1}
\cot (\alpha /2) \\
&=& -\frac{2^N }{(N-1)!} \left( \frac{d}{d\alpha } \right)^{N-1}
\sum_{k=1}^{\infty } \sin k\alpha \\
&=& -\frac{2^N }{(N-1)!} \left( \frac{d}{d\alpha } \right)^{N-1}
\frac{1}{2i} \sum_{k\neq 0} \mbox{sgn} (k) e^{ik\alpha } \\
&=& \frac{1}{2} \sum_{k\neq 0} \frac{1}{|k|} e^{ik\alpha }
\frac{1}{(N-1)!} (2ik)^N
\label{tdern}
\end{eqnarray}
Note that the manipulations of Fourier series carried out above are
well-defined in the context of the theory of distributions
\cite{light}.

As long as one considers continuous momenta, many quantities
connected with the plasmon contribution to the partition function can
be evaluated explicitely even for finite parameter $\alpha $, in analogy
with the ideal gas expressions given in equations (\ref{frklnr}) and
(\ref{frm0}). Thus, in the classical nonrelativistic limit one has
\begin{equation}
G_{12} = \frac{\beta }{8\pi } \sqrt{ \frac{2m\pi }{\beta } }
e^{-\beta (m-\mu /2 )} , \ \ \ \ \ \mbox{Re} \,
G_{11} = \cos (\alpha /2) G_{12}
\end{equation}
and therefore, using (\ref{qcon}),
\begin{eqnarray}
\ln Z_R (\alpha ) &=& -\frac{gL}{4}
\left( \frac{2m\beta }{\pi } \right)^{1/4}
e^{-\frac{\beta }{2} (m-\mu /2)} \left[ \sqrt{\cos (\alpha /2)} \,
\theta (\cos (\alpha /2)) \right. \nonumber \\
& & \ \ \ \ \ \ \ \ \ \ \ \ \ \ \ \ \ \ \ \ \ \ \ \ \ \ \left.
+ \sqrt{\frac{8}{\pi } } \sqrt{\sin (\alpha /2)}
\sum_{k=1}^{\infty } \frac{\sin (k\alpha + \pi /4)}{\sqrt{k} } \right]
\label{disc}
\end{eqnarray}
where $\theta (x)$ denotes the step function.
Note that the $k$-sum can be expressed as a generalized zeta function
\cite{magnus}. Sinilarly, in the limit $m=\mu =0$ one obtains
\begin{equation}
G_{12} = \frac{1}{4\pi } \frac{\alpha /2 }{\sin (\alpha /2)} , \ \ \ \ \
\mbox{Re} \, G_{11} = \frac{1}{4\pi }
\end{equation}
and from this
\begin{equation}
\ln Z_R (\alpha ) = -\frac{gL}{\sqrt{8\pi } } \left[ 1 +
\sqrt{ \frac{4\alpha }{\pi } } \sum_{k=1}^{\infty }
\frac{\sin (k\alpha + \pi /4)}{\sqrt{k} } \right]
\end{equation}
\end{appendix}

\begin{figure}
\caption{Ideal gas part of the logarithm of the partition function
in the ultrarelativistic limit $m=\mu =0$. Short-dash-dotted: Grand
canonical ensemble with discrete momenta (periodic boundary
conditions); long-dash-dotted: Same, with antiperiodic boundary
conditions. Solid line: Canonical ensemble with continuous momenta;
short dashes and long dashes: Same with discretized momenta, periodic
and antiperiodic boundary conditions, respectively. A grand canonical
calculation with continuous momenta gives a constant, namely the one
all curves converge to at large L.}
\label{fig1}
\end{figure}

\begin{figure}
\caption{Ideal gas part of the logarithm of the partition function
in the classical nonrelativistic limit: $m/T=10$, $\mu /T =14$.
Short-dash-dotted: Grand canonical ensemble with discrete momenta
(periodic boundary conditions); long-dash-dotted: Same, with
antiperiodic boundary conditions. Solid line: Canonical ensemble;
in this case the curves obtained with continuous momenta and with
momenta discretized in the various ways are indistinguishable. A
grand canonical calculation with continuous momenta gives a constant,
namely the one all curves converge to at large L.}
\label{fig2}
\end{figure}

\begin{figure}
\caption{Coulomb interaction}
\label{vertex}
\end{figure}

\begin{figure}
\caption{Sum of ring diagrams}
\label{ringsum}
\end{figure}

\begin{figure}
\caption{Plasmon contribution to the logarithm of the partition function
in the ultrarelativistic limit $m=\mu =0$, with $g/T =0.1$. Dotted:
Grand canonical ensemble with continuous momenta; solid line: Same,
with discretized momenta. In the latter case, periodic and antiperiodic
boundary conditions for the fermions give the same, shown, curve. Long
dashes: Canonical ensemble with continuous momenta; short dashes: Same
with discretized momenta. Again, the boundary conditions chosen for the
fermions do not make a difference.}
\label{fig5}
\end{figure}

\begin{figure}
\caption{Plasmon contribution to the logarithm of the partition function
in the classical nonrelativistic limit: $m/T =10, \mu /T =14 $, with
$g/T =0.1$. Dotted: Grand canonical ensemble with continuous momenta; solid
line: Same, with discretized momenta. In the latter case, periodic and
antiperiodic boundary conditions for the fermions give the same, shown,
curve. Long dashes: Canonical ensemble with continuous momenta; short
dashes: Same with discretized momenta. Again, the boundary conditions
chosen for the fermions do not make a difference.}
\label{fig6}
\end{figure}

\end{document}